\documentclass[3p,times,procedia]{elsarticle}
\usepackage{nupha_ecrc}

\volume{00}

\firstpage{1}

\journalname{Nuclear Physics A}

\runauth{S. Hauksson et al.}

\jid{nupha}

\jnltitlelogo{Nuclear Physics A}

\usepackage{amssymb}

\usepackage[figuresright]{rotating}

\usepackage{subcaption}
\usepackage{amsmath}

\newcommand{\beq}{\begin{equation}}
\newcommand{\eeq}{\end{equation}}
\newcommand{\ret}{\mathrm{ret}}
\newcommand{\adv}{\mathrm{adv}}

\begin{document}

\begin{frontmatter}



\dochead{XXVIIIth International Conference on Ultrarelativistic Nucleus-Nucleus Collisions\\ (Quark Matter 2019)}

\title{Hard probes of non-equilibrium quark-gluon plasma}


\author[McGill]{Sigtryggur Hauksson}
\author[McGill]{Sangyong Jeon}
\author[McGill]{Charles Gale}

\address[McGill]{Ernest Rutherford Physics Building,
McGill University,
3600 rue University,
Montréal, QC,
Canada H3A 2T8}

\begin{abstract}
Jets and photons could play an important role in finding the transport coefficients of the quark-gluon plasma. To this end we analyze their interaction with a non-equilibrium quark-gluon plasma. Using new field-theoretical tools we derive two-point correlators for the plasma which show how instabilities evolve in time. This allows us, for the first time, to derive finite rates of interaction with the medium. We furthermore show that coherent, long-wavelength instability fields in the Abelian limit do not modify the rate of photon emission or jet-medium interaction.
\end{abstract}


\end{frontmatter}


\section{Introduction}

Heavy-ion collisions at RHIC and LHC produce the quark-gluon plasma (QGP), a relativistic fluid governed by the strong interaction. These experiments have shown that the QGP is a nearly perfect fluid with the lowest ratio of shear viscosity to entropy density of any known material. A primary goal is to quantify the value of the shear viscosity and other transport coefficients of the QGP.
Soft hadrons are the experimental probe most often used for this purpose. They are produced when the QGP has expanded and cooled sufficiently for the quarks and gluons to coalesce into hadrons and thus they are mostly sensitive to the last stages of the collisions. Penetrating probes, such as jets and photons, are sensitive to a larger part of the evolution of the QGP and could thus be used to extract transport coefficients accurately. Doing so requires a detailed  understanding of non-equilibrium QGP and how these probes interact with it. Such an understanding has been hindered by various theoretical issues such as plasma instabilities, which we discuss here. 

A calculation of hard probes in a non-equilibrium plasma starts with a description of the plasma itself. A weakly coupled QGP is described by two energy scales given certain conditions \cite{AMYkinetic}. Firstly, there are quark and gluon quasiparticles with energy \(\Lambda\). Because they are localized and only interact occasionally, they can be described with a Boltzmann equation 
\beq
\label{Eq:Boltzmann}
\frac{\partial f}{\partial t}  + \mathbf{v} \cdot \frac{\partial f}{\partial \mathbf{r}} + \mathbf{F} \cdot \frac{\partial f}{\partial \mathbf{p}} = \mathcal{C}[f,A].
\eeq
Here \(f\) is the occupation density of the quasiparticles and interaction between them is described by a collision kernel \(\mathcal{C}\).
Between interaction the quasiparticles are deflected in a gluonic, long-wavelength background field as described by the external force \(F\).  Secondly, the gluonic background fields have energy \(g\Lambda\) where \(g \ll 1\) is the coupling constant. Their occupation density is high, which means that they obey the classical equations of motion for non-Abelian fields,
\beq
\label{Eq:YM}
\mathcal{D}_{\mu} F^{\mu\nu} = j^{\nu}
\eeq
where \(F^{\mu\nu}\) is the field strength tensor, $\mathcal{D}$ is a covariant derivative and the current \(j\) comes from quasiparticles. These two equations are coupled as the background field deflects the quasiparticles and the quasiparticles source the background field.

\begin{figure}
\begin{center}
    \begin{subfigure}[b]{0.13\textwidth}
        \centering
        \includegraphics[width=\textwidth]{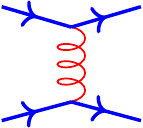}
        \caption{Two-to-two scattering of quasiparticles}
        \label{Fig:2to2}
    \end{subfigure}
    \quad\quad\quad
    \begin{subfigure}[b]{0.28\textwidth}
        \centering
        \includegraphics[width=\textwidth]{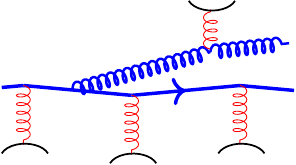}
        \caption{Medium-induced splitting}
        \label{Fig:jet-splitting}
    \end{subfigure}
    \quad\quad\quad
         \begin{subfigure}[b]{0.28\textwidth}
         \centering
        \includegraphics[width=\textwidth]{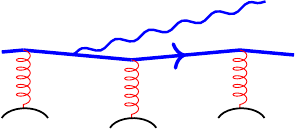}
        \caption{Medium-induced photon emission}
        \label{Fig:photon-splitting}
    \end{subfigure}
    \caption{Interaction of quasiparticles in a weakly coupled quark-gluon plasma}
\end{center}
\end{figure}

The interaction between quark and gluon quasiparticles comes from two equally important processes, two-to-two scattering as seen in Fig. \ref{Fig:2to2} and medium-induced bremsstrahlung as seen in Fig. \ref{Fig:jet-splitting}. This latter process will be our focus. Its rate depends on the strength of the long-wavelength background fields. An on-shell particle radiates an on-shell gluon while receiving gentle kicks from the background fields which bring it slightly off shell. These kicks act coherently and need to be summed up to calculate the rate at leading order in the coupling constant. The total result of the coherent kicks is to reduce the rate of splitting; this is known as the Landau-Pomeranchuk-Migdal (LPM) effect. Nearly identical processes govern jet-medium interactions where the emitter is an energetic jet particle, and photon emission seen in Fig. \ref{Fig:photon-splitting} \cite{AMYgluon}.

\section{Correlators in an unstable plasma} 

Our goal is to evaluate the rate of jet-medium interaction or photon emission through medium-induced bremsstrahlung in a non-equilibrium medium. Unfortunately, one arrives at a divergent rate in a naive calculation that simply swaps equilibrium and non-equilibrium quasiparticle momentum distributions. At a purely mathematical level the divergence arises because of  poles \(i\gamma\)  in the retarded propagator of long-wavelength gluons. This can be seen by solving Eqs. \eqref{Eq:Boltzmann} and \eqref{Eq:YM} together to eliminate the hard particles \cite{Romatschke2003}. In the time domain such a pole yields  exponentially growing modes \(e^{\gamma t}\) with growth rate \(\gamma \).
The physics of these rapidly growing modes is well known \cite{Mrowczynski2016}. They arise for any initial momentum distribution of quark and gluon quasiparticles which is not isotropic in momentum, \(f_0(\mathbf{p}) \neq f_0(p)\). In fact the growth rate is \(\gamma \sim \xi g \Lambda\) where \(\xi\) is the anisotropy of the momentum distribution.  The quarks and gluons radiate long-wavelength chromomagnetic fields which in turn deflect them. It turns out that this increases the anisotropy of the momentum distribution and thus enhances the radiation of chromomagnetic fields even further. This is known as Weibel instabilites. Thus the field strength grows exponentially until more complicated physics such as non-Abelian interaction of the chromomagnetic fields starts to dominate. 

Using a simple analogy with non-relativistic quantum mechanics, it is easy to see how the exponentially growing modes lead to seemingly divergent rates for medium-induced bremsstrahlung. We consider a two-level system with energy difference \(\omega_0\) in an external potential \(V(\mathbf{r}) e^{\gamma t}\), the strength of which grows exponentially. If we turn the potential on at time \(t_0=-\infty\) the transition probability between the levels goes like
\(\left|e^{(i\omega_0 + \gamma)t}/(i\omega_0 + \gamma)\right|^2\) as seen using time-dependent perturbation theory. This diverges in the limit \(\omega_0, \gamma \rightarrow 0\) where the external potential has the same strength for an infinite time and the energy difference between the two levels vanishes. However, if we turn the potential on at time \(t_0=0\) the transition probability goes like 
\(\left|(e^{(i\omega_0 + \gamma)t}-1)/(i\omega_0 + \gamma)\right|^2\) which is perfectly finite in the same limit. This translates directly to our case of medium-induced bremsstrahlung where \(\gamma\) is the growth rate of instabilities and \(\omega_0\) is the energy of kicks the emitting particle receives from background fields.  Thus it is clear that in order to get a finite rate for medium-induced bremsstrahlung in a non-equilibrium medium we need to take into account the time evolution of the background fields and the finite initial time \(t_0\).

The time evolution of the long-wavelength background fields is best described by two-point correlators. To derive the correlators we assume that \(\xi \ll g\) so that the instability mode  grows slower than the typical time of medium-induced bremsstrahlung \(1/g^2\Lambda\). This ensures a kinetic theory description of quasiparticles where interaction can be treated as instantaneous. The requirement \(\xi \ll g\) means that we start with a slightly anisotropic momentum distribution of quasiparticles at time \(t_0 = 0\). The retarded propagator for long-wavelength gluons can then be written as
 \beq
G_{\ret}(k^0) = \widehat{G}_{\ret}(k^0) + \sum_{i} \frac{B}{k^0 - i\gamma}
\eeq
where \(\widehat{G}_{\ret}\) has poles of order \(g\Lambda\) and describes fluctuations emitted by the quasiparticles while the second term describes exponential growth of the instability modes. 

The density of these different modes is given by the \(rr\) correlator \(G_{rr}(x,y) = \frac{1}{2} \langle \left\{ A(x),A(y)\right\}\rangle\). It is the crucial ingredient to calculate medium-induced bremsstrahlung where the rate of kicks depends directly on the density of the background field. 
Using tools of non-equilibrium quantum field theory we obtain \cite{inprep}
\beq
\label{Eq:rr}
\begin{split}
G_{rr}(t_x,t_y) \approx
\int \frac{dk^0}{2\pi} \;e^{-ik^0(t_x-t_y)}\;\widehat{G}_{\ret}(k^0) \,\Pi_{aa}(k^0)\, \widehat{G}_{\adv}(k^0) \\
+ \;\frac{B \,\Pi_{aa}(0) B^{\dagger}}{2\gamma} \left[ e^{\gamma t_x} e^{\gamma t_y}-1\right]
\end{split}
\eeq  
The first term has poles at energy scale \(g\Lambda\) and has the same form as the equilibrium result except that all instability poles have been subtracted in \(\widehat{G}_{\ret}\) and \(\widehat{G}_{\adv}\). The second term is at energy scale \(\gamma \sim \xi g \Lambda\) and clearly shows the exponential growth. The important point is that because of the initial time we get 
\beq
\frac{1}{2\gamma} \left[ e^{\gamma t_x} e^{\gamma t_y}-e^{\gamma t_x} e^{\gamma t_y} \bigg\rvert_{t_x, t_y = 0}\right]
\eeq
just as in the quantum mechanical example above. This expression is perfectly finite as \(\gamma \rightarrow 0\) which ensures that the rate of medium-induced bremsstrahlung is finite.

\section{Medium-induced bremsstrahlung in a non-equilibrium medium}

We now explore how the background fields at energy \(\xi g\Lambda\) can change the physics of medium-induced bremsstrahlung. To fix ideas, we focus on photon emission as in Fig. \ref{Fig:photon-splitting} but our results can be extended easily to medium-induced splitting as in Fig. \ref{Fig:jet-splitting}. Our calculation is valid for any \(rr\) propagator that has two scales: a fluctuating field at energy \(g\Lambda\) and a coherent background field at much lower energy \(\xi g \Lambda\). 

\begin{figure}
        \centering
        \includegraphics[width=0.40\textwidth]{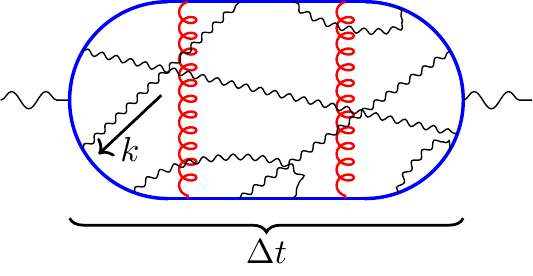}
    	\caption{Diagrams for medium-induced bremsstrahlung of a photon. Interaction with medium is composed of time-ordered \(g\Lambda\) kicks (red) and kicks from a coherent field at energy \(\xi g \Lambda\) (black). The latter are not ordered in time. Here \(\Delta t\) is the time emission takes and \(k \sim \xi g \Lambda\) is the momentum flow in the coherent fields.}
    	\label{Fig:diagram}
\end{figure}

The effect of the coherent background field at energy \(\xi g\Lambda\) is very diffent from the effect of the fluctuating field at energy \(g\Lambda\). Since bremsstrahlung takes time \(1/g^2\Lambda\) and kicks from the fluctuating field take time \(1/g\Lambda\) there is no overlap between subsequent kicks: they are ordered in time. In diagrammatic language this means that diagrams with crossed gluon rungs are suppressed. Interaction with the coherent background field, on the other hand, takes much longer time \(1/\xi g\Lambda\) so that there is no time ordering. Diagrammatically, this means that diagrams with crossed rungs need to be resummed, see Fig. \ref{Fig:diagram}. Because of interaction with the coherent background field the diagrams have complicated color factors and thus are only amenable to evaluation in the Abelian limit or in the planar, large \(N_c\) limit. In these proceedings we  focus on the Abelian case.

We will use approximations to sum up the kicks from the Abelian, coherent field. We expand in the small parameter \(k\Delta t \sim \xi/g \ll 1\) where \(k\) is the momentum of the coherent field and \(\Delta t\) is the time for the emission of a photon. After resummation, the leading order contribution in \(k \Delta t\) can be shown to be of the form \(\sim e^{-\int_k G_{rr} (\Delta t)^2}\) which corresponds to phase rotation of the emitting quark in the background field. The next-to-leading order contribution goes like \(\sim e^{-\int_k G_{rr} k^2 (\Delta t)^4}\). The factor of \(k^2\) in the exponential can be seen as a double derivative acting on \(G_{rr} \sim A^2\) and thus gives the field strength \(F^{\mu\nu}\). This means that the next-to-leading order correction describes a change in the dispersion relation of the quark due to the background electromagnetic field. At even higher order there are other corrections, such as spin procession and rotation of momentum distribution of the quarks. We do not consider these here.

Remarkably, the effect of the coherent background Abelian field vanishes at leading and next-to-leading order in \(k\Delta t\) \cite{inprep}. 
A similar cancellation takes place in jets in vacuum where IR divergences due to very soft emission are cancelled by virtual effects. However, it is unexpected that this cancellation takes place even when \(g\Lambda\) kicks are included. This cancellation does not take place in the large \(N_c\) limit which will be the topic of future work.

There are important non-equilibrium effects in medium-induced bremsstrahlung even though the coherent background field makes no difference in the Abelian case. These non-equilibrium effects come from resummation of the \(g\Lambda\) kicks which are different from the equilibrium case. A detailed analysis \cite{Hauksson2017} shows that the rate of photon emission goes like \(\mathrm{Re}\, \mathbf{f}(\mathbf{p}_{\perp})\) which can be understood as the probability for the quark to gain transverse momentum \(\mathbf{p}_{\perp}\) because of \(g\Lambda\) kicks from the fluctuating fields. This probability is determined by a Boltzmann-like equation
\beq 
 \mathbf{p_{\perp}} = i \delta E\; \mathbf{f}(\mathbf{p_{\perp}}) + \int_{\mathbf{q_{\perp}}} \; \mathcal{C}(\mathbf{q}_{\perp}) \left[\mathbf{f}(\mathbf{p_{\perp}}) - \mathbf{f}(\mathbf{p_{\perp}} + \mathbf{q}_{\perp})\right].
\eeq
where the collision kernel
\beq
\mathcal{C} \sim \int_{q^0}  \widehat{G}_{\ret} \Pi_{aa} \widehat{G}_{\adv} \rvert_{q^z = q^0} 
\eeq
determines the rate of interaction with the fluctuating fields. Importantly, our derivation shows that one should subtract instability poles from the retarded and advanced function in this expression. 
This prescription thus allows for the evaluation of jet-medium interaction and photon emission in a non-equilibrium QGP \cite{Hauksson2018}, and the use of these probes to extract transport coefficients.\\

 {\noindent }\textbf{Acknowledgments:} This work was funded in part by the Natural Sciences and Engineering Research Council of Canada, and by the Fonds de recherche du Qu\'ebec -- Nature et technologies.





\bibliographystyle{elsarticle-num}
\bibliography{bibliography.bib}







\end{document}